\documentclass[conference,letterpaper,twoside,twocolumn]{IEEEtran}
\usepackage{amsmath,amssymb,amscd,latexsym,dsfont}
\usepackage{color}
\usepackage{comment}
\usepackage{enumerate}
\usepackage{cite}
\usepackage{amsmath}
\usepackage{mathrsfs}
\usepackage{graphicx}
\usepackage{psfrag}
\usepackage{footnote}
\usepackage[usenames,dvipsnames]{xcolor}
\DeclareMathOperator{\sech}{sech}

\IEEEoverridecommandlockouts

\newtheorem{theorem}{Theorem}

\newtheorem{lemma}{Lemma}

\title{A Lower Bound on the per Soliton Capacity of the Nonlinear Optical Fibre Channel}
\author{
    \IEEEauthorblockN{Nikita A. Shevchenko\IEEEauthorrefmark{1}, Jaroslaw E. Prilepsky\IEEEauthorrefmark{2}, Stanislav A. Derevyanko\IEEEauthorrefmark{4}, \\
		 Alex Alvarado\IEEEauthorrefmark{1}, Polina Bayvel\IEEEauthorrefmark{1} and Sergei K. Turitsyn\IEEEauthorrefmark{2}}\\
		\IEEEauthorblockA{\IEEEauthorrefmark{1}Optical Networks Group, University College London, Torrington Place, London WC1E 7JE, United Kingdom\\}
    \IEEEauthorblockA{\IEEEauthorrefmark{2}Aston Institute of Photonic Technologies, Aston University, Birmingham B4 7ET, United Kingdom}
		\IEEEauthorblockA{\IEEEauthorrefmark{4}Department of Physics of Complex Systems, Weizmann Institute of Science, Rehovot 76100, Israel}		
		\thanks{Research supported by the Engineering and Physical Sciences Research Council (EPSRC) project UNLOC (EP/J017582/1), United Kingdom.}
}

\begin{document}
\maketitle

\begin{abstract}
A closed-form expression for a lower bound on the per soliton capacity of the nonlinear optical fibre channel in the presence of (optical) amplifier spontaneous emission (ASE) noise is derived. This bound is based on a non-Gaussian conditional probability density function for the soliton amplitude jitter induced by the ASE noise and is proven to grow logarithmically as the signal-to-noise ratio increases.
\end{abstract}

\section{Introduction}
It is widely accepted that in order to meet the ever-growing demand for data rates in fibre-optic telecommunication systems, the spectral efficiency of the optical fibre transmission system needs to be increased \cite{Richardson}. The key physical effects distinguishing a fibre optical system from a free space transmission are: dispersion, nonlinearity and optical noise \cite{Essiambre,Hasegawa,Iannone,Mollenauer}. The implementation of the ``fifth generation'' of optical transceivers and networks operating with coherent detection, advanced multilevel modulation formats, and digital signal processing techniques, has led to the possibility of channel rates exceeding 100 Gbit/s \cite{cd13}. The key to this breakthrough is the mitigation of linear transmission impairments, such as chromatic and polarization mode dispersion.

The performance of current coherent systems is limited by noise and nonlinearity. In contrast to linear channels, however, spectral efficiencies for the optical fibre channel usually exhibit a peak and decay at high input powers; this is often referred to as the ``nonlinear Shannon limit'' \cite{Ellis,Mecozzi}. This behaviour is caused by the Kerr nonlinearity and is believed to ultimately lead to a ``capacity crunch'' \cite{Richardson}, i.e., to the inability of the optical network infrastructure to cope with the increasing capacity demand.

The capacity analysis of the nonlinear channel relies on well-established methods of information theory \cite{Shannon,CoverThomas}. However, most of the analytical results obtained to date concern linear channel models, and hence, are not directly applicable to nonlinear optical channels. Despite numerous efforts to define the influence of Kerr nonlinearity on the channel capacity \cite{Splett,Mitra,Stark,Wegener,Ellis,Mecozzi,Secondini,Dar}, the capacity of the nonlinear optical channel still remains as an open research problem. Most of the capacity bounds presented in the literature typically display a peaky behaviour, where the maximum is reached at a finite threshold power. To the best of our knowledge, the first nondecaying (lower) bound on the capacity of the nonlinear optical fibre channel (with zero average dispersion) was presented in \cite{tdy03}. Other nondecaying bounds include, e.g., those given recently in \cite{Agrell14} and \cite{Yousefi15,Kramer15}.

A multitude of different nonlinearity mitigation techniques have been proposed over recent years to suppress nonlinearity-induced distortions. This includes receiver-based digital signal processing \cite{Millar}, digital back-propagation \cite{Ip}, optical phase conjugation \cite{Du}, twin-waves phase conjugation \cite{Liu}, etc. However, there are still many limitations and further challenges in applying these methods. A promising alternative for nonlinearity compensation is the nonlinear Fourier transform (NLFT) developed in the 70's \cite{Zakharov,AKNS}. The applications of the NLFT in optical communication originates from the pioneering work of Hasegawa and Nyu \cite{Nyu}, an approach that has been extended in a number of recent works \cite{EGT,Yousefi_1,Yousefi_2,Yousefi_3,Hari,Meron,Prilepsky0,Prilepsky,Le,Le2}. Notably, an experimental demonstration of a NLFT-based transmission was recently presented by B\"ulow in  \cite{Bulow}.

The use of NLFT for nonlinearity compensation in optical fibre links is possible because the master model governing signal propagation in a single mode optical fibre (in the absence of noise and loss) is the nonlinear Schr\"odinger equation (NLSE) \cite{Hasegawa,Iannone,Mollenauer} that belongs to the class of  \textit{integrable} (i.e., completely solvable) evolutionary equations \cite{Zakharov}. The solution method can be considered as the generalisation of the linear Fourier transform (FT) operation onto the nonlinear (integrable) system, hence the name NLFT. Similarly to the FT, the NLFT decomposes a waveform in the NLSE space-time domain into the nonlinear normal modes inside the nonlinear spectral domain \cite{Yousefi_1,Prilepsky}. The key underlying feature of the NLFT transmission is that these nonlinear modes (nonlinear signal spectrum) propagate without cross-talk, effectively in a linear manner. Thus, the nonlinear spectrum can be used for encoding and efficient transmission of information over the nonlinear fibre.

The original work by Hasegawa and Nyu \cite{Nyu} introduced the concept of ``eigenvalue communications'', where the information was encoded using discrete eigenvalues associated with the solitonic degrees of freedom emerging from the NLFT signal decomposition \cite{Hasegawa} (see also \cite{Yousefi_1}). In the absence of both loss and noise, the evolution of nonlinear modes is inherently free from any nonlinear impairments, including a nonlinear cross-talk. The loss in optical links is usually compensated by using lumped or distributed amplification; in our case we assume the ideal distributed Raman amplification scheme resulting in the lossless NLSE \cite{QL1,QL2,Essiambre}. However, the signal will still be distorted by amplifier-induced spontaneous emission (ASE) as well as signal-noise beating.

In this paper, we study the channel capacity (in bits per soliton symbol)\footnote{The practically more relevant problem of channel capacity in [bit/s/Hz] is left for future investigation.} for a transmission system based on optical solitons (sufficiently separated in time domain) launched into a noisy NLSE channel. The information is assumed to be encoded in the soliton's amplitude only, which can be extracted from the imaginary part of the discrete eigenvalue emerging from the NLFT signal decomposition. We consider a discrete-time continuous-input continuous-output channel model, based on the asymptotically exact non-Gaussian marginal statistics of the soliton amplitude in the presence of weak ASE noise presented in \cite{FLKT,Derevyanko,dty05}. We emphasise that the capacity estimations for such fundamentally nonlinear channels are quite few and far between. Notable exceptions are the works by Yousefi and Kschischang \cite{Yousefi_3} and Meron et al. \cite{Meron}. While in \cite{Yousefi_3} the channel statistics were assumed a priori to be Gaussian \cite[eq.~(27)]{Yousefi_3}, in \cite{Meron} a tight lower bound on the channel capacity as a function of the signal to noise ratio (SNR) was not provided.

The discrete-time channel model governing transmission systems based on optical solitons is a noncentral chi-squared distribution with four degrees of freedom \cite{Derevyanko,dty05}. Based on this model we obtain an asymptotically growing lower bound for the channel capacity vs. SNR. This bound is similar to the one in \cite{tdy03}, where a noisy nonlinear optical fibre channel with zero fibre dispersion was considered. The results in this paper show that the reachable capacity limits for existing optical fibre channels could have been previously underestimated.

\section{The Master Equation and the Noncentral Chi-squared Channel Model}

\subsection{Waveform channel}

We consider propagation of a slowly varying envelope signal formed by a sequence of solitons transmitted every $T_{s}$~[s] over a nonlinear optical fibre. Our model combines the effects of chromatic dispersion (we consider the case of anomalous dispersion), instantaneous Kerr  nonlinearity, and ASE noise due to optical Raman amplification. The fibre loss is assumed to be continuously compensated along the fibre by means of ideal Raman amplification and hence is set to zero \cite{QL1,QL2,Essiambre}. 

\begin{figure}[tpb]
 \begin{center}
   \newcommand{\scale}{0.65}
   \psfrag{aa}[cc][cc][\scale]{(a)}
   \psfrag{bb}[cc][cc][\scale]{(b)}
	 \psfrag{a}[cc][cc][\scale]{$\frac{\{\cdot\}}{\sqrt{2}}$}
   \psfrag{b}[cc][cc][\scale]{$\{\cdot\}^2$}
   \psfrag{c}[cc][cc][\scale]{$\sqrt{\frac{\{\cdot\}}{2}}$}
   \psfrag{tx}[cl][cl][\scale]{\textbf{Transmitter}}
   \psfrag{of}[cl][cl][\scale]{\textbf{Optical Fibre}}
   \psfrag{rx}[cl][cl][\scale]{\textbf{Receiver}}
	 \psfrag{wfch}[cc][cc][0.7]{\textit{Waveform channel}}
	 \psfrag{dtch}[cc][cc][0.7]{\textit{Discrete-time channel}}
   \psfrag{xk}[cc][cc][\scale]{$X_k$}
   \psfrag{yk}[cc][cc][\scale]{$Y_k$}
   \psfrag{st}[cc][cc][\scale]{$s(t)$}
   \psfrag{rt}[cc][cc][\scale]{$r(t)$}
   \psfrag{ase}[cc][cc][0.5]{ASE noise}
	 \psfrag{asm}[cc][cc][\scale]{{\parbox[b]{10em}{\centering Amplitude-\\to-soliton\\Mapper}}}
   \psfrag{sam}[cc][cc][\scale]{{\parbox[b]{10em}{\centering forward\\NLFT}}}
   \psfrag{noise1}[cl][cl][0.6]{$N_{1} \sim \mathcal {N} \big(0,\sigma_\mathrm{N}^2 \big)$}
	 \psfrag{noise2}[cl][cl][0.6]{$N_{2} \sim \mathcal {N} \big(0,\sigma_\mathrm{N}^2 \big)$}
	 \psfrag{noise3}[cl][cl][0.6]{$N_{3} \sim \mathcal {N} \big(0,\sigma_\mathrm{N}^2 \big)$}
	 \psfrag{noise4}[cl][cl][0.6]{$N_{4} \sim \mathcal {N} \big(0,\sigma_\mathrm{N}^2 \big)$}
   \centering
	 \includegraphics[scale=0.62]{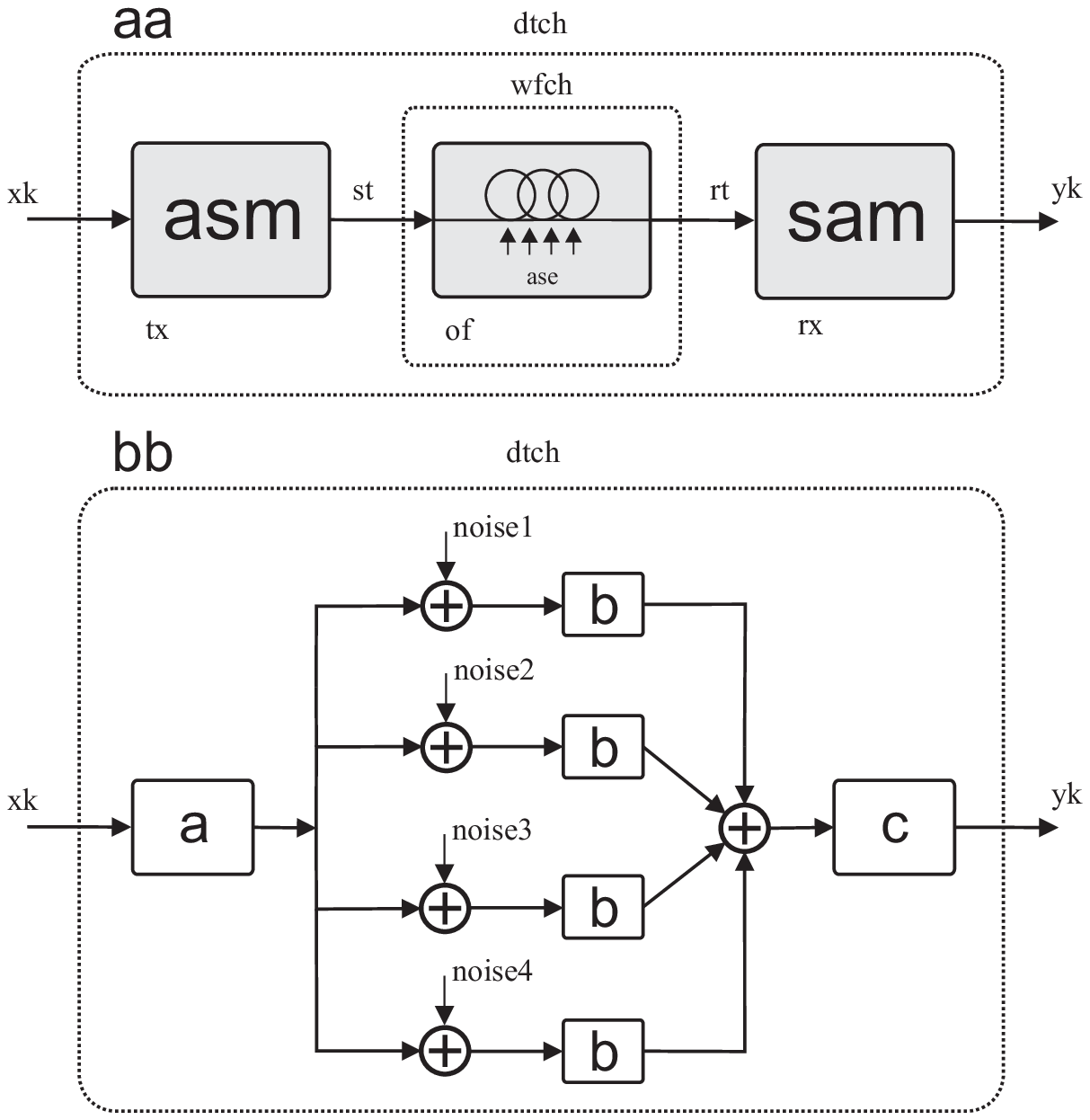}
	 \vspace{-.4cm}
  \end{center}
 \caption{\label{System model} System model: (a) Transmitter, continuous-channel model governed by \eqref{NLSE}, and receiver. (b) Equivalent discrete-time channel model.}
 \vspace{-0.4cm}
\end{figure}

The noise-perturbed NLSE in dimensionless units is given by \cite{Essiambre,Hasegawa,Derevyanko}
\begin{equation}
i \frac{\partial q(z,t)}{\partial z} + \frac{1}{2} \frac{\partial^2 q(z,t)}{\partial t^2} + |q(z,t)|^2 q(z,t) = n(z,t),
\label{NLSE}
\end{equation}
where $t$ is the time normalised by $T_{s}$, $z$ is the distance along the fibre normalised by $L_{s} \triangleq \frac{T_{s}^2}{|\beta_2|}$ (not to be confused with the dispersion length), and $\beta_2<0$ is the group velocity dispersion coefficient. We also define $s(t)=q(0,t)$ and $r(t)=q(L,t)$ as the input and output waveforms of the physical channel after transmission distance $L$, respectively, normalised by the nonlinear power scale $(\gamma L_{s})^{-1}$, where $\gamma$ is the nonlinearity coefficient. The relationship between $s(t)$ and $r(t)$ is schematically shown in the inner part of Fig.~\ref{System model}(a).

The noise term $n(t,z)$ on the right-hand side of (\ref{NLSE}) is assumed to be a zero-mean ($\mathbb {E} \, [ n(z,t) ] = 0$) circularly-symmetric additive white Gaussian noise (AWGN) process with autocorrelation function \cite[eq.~(53)]{Essiambre}
\begin{equation}
\label{noise}
\mathbb {E} \, [ n(z,t) \, n^*(z',t') ] = 2 D \, \delta(z-z') \, \delta(t-t'),
\end{equation}
where $^*$ denotes complex conjugation, $\mathbb {E} \, [\cdot]$ is the mathematical expectation operator, and $\delta(\cdot)$ is the Dirac's delta function. Here the noise intensity $D$ is written in dimensionless units \cite[eq.~(5.29)]{Iannone} as $D = \gamma L_{s}^2 \, \sigma_{0}^2/2 T_s$. For ideal distributed Raman amplification, the power spectral density of the ASE noise $\sigma_{0}^2$ is defined as \cite[eq.~(56)]{Essiambre} $\sigma_{0}^2 =  \alpha  K_{T} \cdot h\nu_{\mathrm{opt}}$, where $\alpha$ is the fibre attenuation coefficient, $h\nu_{\mathrm{opt}}$ is the average photon energy, and $K_{T} \approx 1$ is the coefficient that characterizes the Raman pump providing the gain \cite{Essiambre}.

It is known that the noiseless NLSE (i.e., \eqref{NLSE} with $n(z,t)=0$) possesses a special class of solutions, the so-called fundamental bright solitons \cite{Hasegawa,Iannone,Mollenauer}. At $z=0$ we write it as \cite[eq.~(1.40)]{Mollenauer} (in normalised units)
\begin{equation}
q(0,t) = A_{0} \sech (A_{0} t),
\label{soliton.1}
\end{equation}
where $A_{0}$ denotes the normalised soliton amplitude and we assume that the initial soliton frequency, phase and centre-of-mass position are set to zero. The unperturbed soliton solution (\ref{soliton.1}) at a distance $z=L$ is given by $q(L,t) = A_{0} \sech (A_{0} t) \exp(i A_0^2 L/2)$.

\subsection{Discrete-time channel}
The (normalised) continuous-time input signal $s(t)$ is defined as
\begin{equation}
s(t) = \sum_{k=1}^\infty s_k(t),
\label{signal}
\end{equation}
where
\begin{equation}
s_k(t) = A_{0 k} \sech [A_{0 k}(t - k)],
\label{signal.1}
\end{equation}
and $k$ is the discrete-time index. At each discrete time $k$, the transmitter maps an amplitude $A_{0 k}$ to $s_k(t)$ via \eqref{signal.1}. For simplicity of the following analysis, however, we consider the square root of the amplitudes, i.e., $X_{k} = \sqrt{A_{0 k}}$.

The dimensionless energy of the $k$th soliton waveform is defined as
\begin{equation}
\label{average_power}
E(A_{0 k}) \triangleq \intop_{-(k-1/2)}^{(k+1/2)} |s_k(t)|^2 \mathrm{d} t.
\end{equation}
We consider the regime where the inter-soliton separation $T_s$ is much larger than the typical soliton width (low duty cycle), so the integral in \eqref{average_power} can be taken over $(-\infty, \infty)$. This yields the well-known linear energy-amplitude scaling of the soliton pulse $E(A_{0 k})= 2 A_{0 k}$. The minimum inter-soliton separation is then determined by the peak power $A_{0k}^2$ of each individual soliton, which is in turn inversely proportional to the square of its width $T_{0k}=\frac{1}{A_{0 k}}$.

The receiver in Fig. 1(b) processes the received waveform $r(t)$ during a window of length one via the forward NLFT and returns the amplitude of the received soliton. We assume ideal NLFT-detection, i.e., the sampling rate is high enough to ignore NLFT finite accuracy issues arising from a particular algorithmic realisation \cite{Yousefi_2} compared to noise-induced distortions. The inter-soliton separation is also assumed to be large enough so that there is no interaction between adjacent solitons, i.e., $\exp(-A_{0k}) \ll 1$, or equivalently, $1 \gg T_{0k}$. Another source of corruption for the soliton-based transmission system emanates from the Gordon-Haus (GH) timing jitter \cite{Mollenauer,Meron}, which defines the standard deviation $\Delta T^{\mathrm{GH}}$ of the soliton position as a function of the propagation distance and soliton amplitude. To avoid interaction between adjacent solitons, the GH timing jitter should also be taken into account\cite{Mollenauer}. For a given propagation distance $L$, the inter-soliton separation must fulfill $1 > T_{0k} + \Delta T^{\mathrm{GH}}$. This condition guarantees that solitons behave as isolated pulses, and thus, there is no intersymbol interference. We assume that this condition is satisfied throughout this paper, and thus, from now on we drop the time index $k$. 

The exact conditional PDF for a single received amplitude $A$ given a transmitted amplitude $A_0$ is written as \cite[eq.~(24)]{Derevyanko} (see also \cite{dty05})
\begin{equation}
p_{A|A_0} (a|a_0) =
\frac{1}{\sigma_\mathrm{N}^2} \, \sqrt{\frac{a}{a_0}} \exp \bigg( - \frac{a_0+a}{\sigma_\mathrm{N}^2} \bigg)
I_1 \bigg( \frac {2 \sqrt{a_0a}}{\sigma_\mathrm{N}^2}\bigg),
\label{PDF}
\end{equation}
where $\sigma_\mathrm{N}^2 = LL_{s}^{-1}\,D/2$ is the normalised variance of the accumulated ASE noise and $I_1(x)$ is the modified Bessel function of the first kind. Expression (\ref{PDF}) is in fact the same PDF obtained assuming an energy-detection receiver (i.e., a receiver based on \eqref{average_power}), as shown in \cite[eq.~(5.501)]{Iannone}.

Equation (\ref{PDF}) is nothing else but a special case of a noncentral chi-squared distribution with four degrees of freedom providing non-Gaussian statistics for soliton amplitudes. For future use, it is convenient to designate the output of the discrete-time channel model $Y$ as \textit{the square root} of the output soliton amplitudes $A$. By making a change of variables, the PDF (\ref{PDF}) can be rewritten as
\begin{equation}
p_{Y|X}(y|x) =
\frac{2}{\sigma_\mathrm{N}^2} \, \frac{y^2}{x}  \exp \bigg( - \frac{x^2+y^2}{\sigma_\mathrm{N}^2} \bigg)
I_1 \bigg( \frac {2 xy}{\sigma_\mathrm{N}^2}\bigg).
\label{originalPDF}
\end{equation}

The conditional PDF in \eqref{originalPDF} describes a channel with the input-output relation
\begin{eqnarray}
Y^2 = \frac{1}{2 } \sum_{i=1}^{4} \, \left(\frac{X}{\sqrt 2} +  N_{i} \right)^2,
\label{channel_model}
\end{eqnarray}
where $N_i$, $i=1,2,3,4$ are four independent and identically distributed zero-mean Gaussian random variables with variance $\sigma_i^2=\sigma_\mathrm{N}^2$. The input-output relationship in \eqref{channel_model} is schematically shown in Fig. 1(b).

\section{Main results}
Since the soliton pulses are assumed to be well separated and the intersymbol interference due to pulse interaction can be neglected, the model (\ref{originalPDF}) describes a scalar memoryless channel. The channel capacity is then defined as \cite{Shannon,CoverThomas}
\begin{equation}
\label{Capacity_def}
C\triangleq \max_{p_{X}}\,I_{XY},
\end{equation}
where $I_{XY}$ is the mutual information (MI) and the optimization is performed over all possible input distributions $p_{X}$ with fixed average symbol energy $\mathbb {E} \, [E(A_{0})]$. The MI $I_{XY}$ can be decomposed as \cite{Shannon,CoverThomas}
\begin{equation}
I_{XY} = h_{Y} - h_{Y|X},
\label{MI}
\end{equation}
where $h_{Y}$ and $h_{Y|X}$ are the output and conditional differential entropies, respectively.

The SNR is defined as \cite[eq.~(29)]{Essiambre}
\begin{equation}
\label{SNR}
\mathrm{SNR} \triangleq \frac{\mathbb {E} \, [ E (A_{0}) ]}{\sigma_\mathrm{N}^2 T_s} = \frac{2 \kappa \sigma_\mathrm{S}^2 }{\sigma_\mathrm{N}^2 },
\end{equation}
where $\sigma_\mathrm{S}^2$ is the average amplitude $\sigma_\mathrm{S}^2=\mathbb {E} \, [ A_0 ] = \mathbb {E} \, [ X^2 ]$ and $\kappa$ is the ratio between the available bandwidth and the symbol rate $1/T_s$. Thus, for a fixed bandwidth and symbol rate, the SNR is proportional to the parameter $\rho \triangleq \sigma_\mathrm{S}^2/\sigma_\mathrm{N}^2$. We shall henceforth consider the capacity and MI as a function of $\rho$.

The exact solution for the power constrained optimization problem (\ref{Capacity_def}) with the channel model (\ref{originalPDF}) is unknown.
To obtain a \textit{lower bound} on the capacity, we shall assume the input symbols $X$ are drawn from a trial input distribution. In this work we use the Rayleigh PDF
\begin{equation}
p_{X}(x) = \frac{2x}{\sigma_\mathrm{S}^2} \exp \bigg( - \frac{x^2}{\sigma_\mathrm{S}^2} \bigg),
\label{Rayleigh}
\end{equation}
which leads to exponentially-distributed soliton amplitudes $A_{0}$ with mean $\sigma_\mathrm{S}^2$.

The next two Lemmas provide exact closed-form expressions for the output differential entropy $h_{Y}$ of symbols $Y$ with input symbols $X$ distributed according to (\ref{Rayleigh}) and for the conditional differential entropy $h_{Y|X}$.
\begin{lemma}
For the channel in (\ref{channel_model}) and the input distribution (\ref{Rayleigh})
\begin{align}
\label{Output Entropy}
\nonumber
h_{Y} =& \log \sqrt{\sigma_\mathrm{S}^2} - \log \sqrt{1+\rho^{-1}} -\rho^{-1} \log \sqrt {1 + \rho} \\
      &+\rho + \psi (\rho^{-1}) - \frac{3}{2} \psi(1)- \log 2 + 1,
\end{align}
where $\psi(x) \triangleq \frac{\mathrm{d}}{\mathrm{d} x} \ln \Gamma(x)$ is the digamma function, and $\Gamma(x)$ is the gamma function.
\end{lemma}
\begin{lemma}
For the channel in (\ref{channel_model}) and the input distribution (\ref{Rayleigh})
\begin{align}
\label{Conditional Entropy}
\nonumber
h_{Y|X} =& \log \sqrt{\sigma_\mathrm{S}^2} + 2 \, ( 1 + \rho) - (1 + \rho^{-1}) \log \, (1 + \rho) \\
        &- \rho^{-1} \sqrt{1+\rho^{-1}} \, F(\rho) - \frac{\psi(1)}{2} - \log 2 ,
\end{align}
where
\begin{align}\label{f}
F(\rho) \triangleq \intop_0^\infty \xi \, K_1 (\sqrt{1+\rho^{-1}} \, \xi) \, I_1(\xi) \log  \big[ I_1(\xi) \big]\mathrm{d} \xi,
\end{align}
and $K_1(x)$ is the modified Bessel function of the second kind of order one.
\end{lemma}
\begin{IEEEproof}[Sketch of the proof]
To prove both lemmas, the output distribution $p_{Y}(y) \triangleq \int_0^\infty p_\mathrm {Y|X}(y|x) \, p_{X}(x) \mathrm{d} x$ is calculated using (\ref{originalPDF}) and (\ref{Rayleigh}). The derived output PDF $p_{Y}(y)$ is then used in the definitions of differential entropies. The results of both Lemmas are then obtained by evaluating the corresponding integrals. The calculation follows closely that from the earlier work \cite{tdy03}, where calculations were performed for a chi-squared distribution with two degrees of freedom (cf. \eqref{f} and \cite[eq.~(24)]{tdy03}).
\end{IEEEproof}

We note that the proof of Lemma~2 includes finding a closed-form expression for the differential entropy of a chi-squared distribution with four degrees of freedom. To the best of our knowledge, this has never been previously reported in the literature.\footnote{However, a closed-form expression for the \emph{expected-log} of a noncentral chi-squared distribution with even number of degrees of freedom was given in \cite[Lemma~10.1]{Lapidoth03}.} The results from Lemmas 1 and 2 can be combined to produce the following theorem.
\begin{theorem}
For the channel (\ref{channel_model}) and the input distribution (\ref{Rayleigh})
\begin{align}
\label{Mutual Information}
\nonumber
I_{XY} =& \log \big( \rho \, \sqrt{1+\rho^{-1}} \, \big) + \rho^{-1} \log \, ( \sqrt{1+\rho} \, ) - \rho \\
       &+ \rho^{-1} \sqrt{1+\rho^{-1}} \, F(\rho) + \psi(\rho^{-1}) - \psi(1) - 1.
\end{align}
\end{theorem}
\begin{IEEEproof}
From Lemmas 1 and 2 and \eqref{MI}.
\end{IEEEproof}

\begin{figure}[t!]
 \begin{center}
   \newcommand{\scale}{0.85}
   \psfrag{hy}[cc][cc][\scale]{$h_{Y}$}
   \psfrag{hyx}[cc][cc][\scale]{$h_{Y|X}$}
	 \psfrag{mi}[cc][cc][\scale]{$I_{XY}$}
	 \psfrag{CF}[cl][cl][0.75]{Closed-form expression}
	 \psfrag{MC}[cl][cl][0.75]{Numerical integration}
	 \psfrag{mibit}[cc][cc][0.95]{Bits per soliton}
	 \psfrag{snr}[cc][cc][0.95]{$\rho$ [dB]}
   \centering
   \includegraphics[scale=0.75]{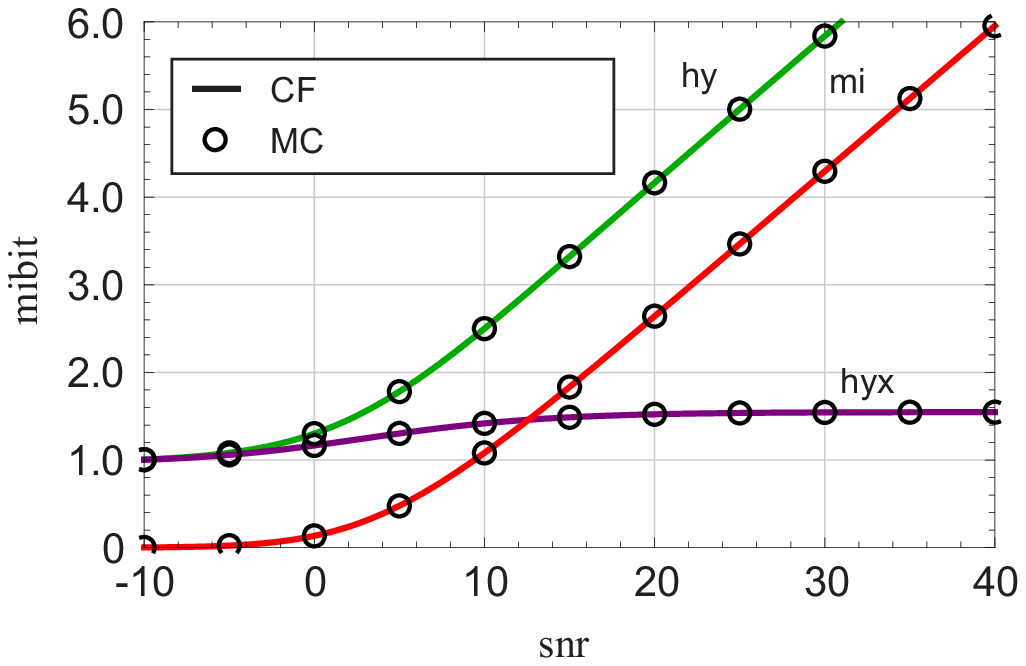}
	\vspace{-.4cm}
 \end{center}
	 \caption{\label{Figure_2} The differential entropy of output distribution $h_{Y}$ in \eqref{Output Entropy}, the differential conditional entropy $h_{Y|X}$ in \eqref{Conditional Entropy}, and the MI $I_{XY}$ in \eqref{Mutual Information}. Results obtained via numerical integration are also shown (circles).}
   \vspace{-0.2cm}
\end{figure}

The results of Lemma 1, Lemma 2, and Theorem 1 are illustrated in Fig.~\ref{Figure_2}. Analytical curves for the functions $h_{Y}$, $h_{Y|X}$, and $I_{XY}$ are compared with results obtained via numerical integration.
\begin{figure}[t!]
 \begin{center}
  \psfrag{formula}[cc][cc][1.00]{$\lim_{\rho \to \infty} \frac{I_{\textnormal{as}}}{I_{XY}} = 1$}
  \psfrag{ratio}[cc][cc][0.95]{$I_{\textnormal{as}}/I_{XY}$}
  \psfrag{snr}[cc][cc][0.95]{$\rho$ [dB]}
  \centering
  \includegraphics[scale=0.74]{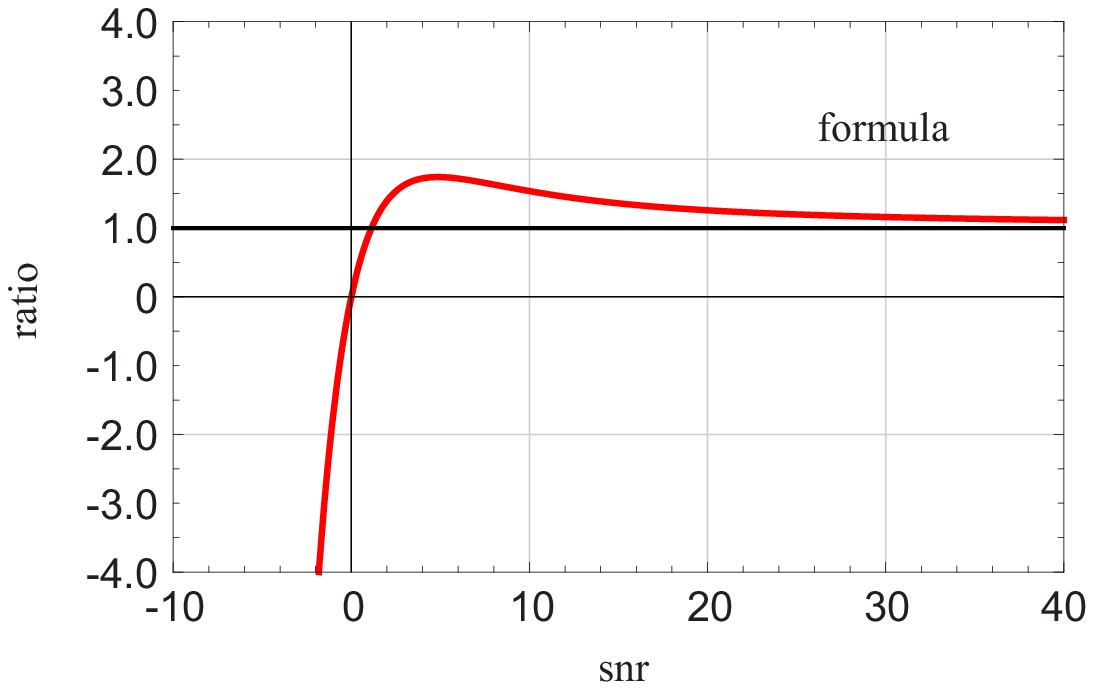}
	\vspace{-.4cm}
 \end{center}
  \caption{\label{Figure_3} The ratio between the MI $I_{XY}$ in \eqref{Mutual Information} and the function $I_{\textnormal{as}}$ in \eqref{a}.}
  \vspace{-0.4cm}
\end{figure}

The next theorem shows that the capacity lower bound is asymptotically equivalent to half the logarithm of the SNR, which is the main result of our work.
\begin{theorem}
The MI $I_{XY}$ in \eqref{Mutual Information} satisfies
\begin{align}
\lim_{\rho \to \infty} \frac{I_{\textnormal{as}}}{I_{XY}} = 1,
\end{align}
where
\begin{align}\label{a}
I_{\textnormal{as}} \triangleq  \frac{1}{2} \log \rho.
\end{align}
\end{theorem}
\begin{IEEEproof}
The proof follows from an asymptotic expansion of $I_{XY}$ in \eqref{Mutual Information} together with the asymptotic expansion of \eqref{f} provided in \cite{tdy03}.
\end{IEEEproof}

Fig.~\ref{Figure_3} shows the numerical evaluation of the ratio $I_{\textnormal{as}}/I_{XY}$ and confirms that the MI behaves asymptotically as $(1/2) \log \rho$, or equivalently, as $(1/2) \log \mathrm{SNR}$. According to Fig.~\ref{Figure_3}, the asymptotic function \eqref{a} approaches the MI from above. Interestingly, the expression \eqref{a} has appeared in asymptotic analyses of optical systems (see e.g., \cite[eq.~(25)]{tdy03}, \cite[Sec.~V-A]{Yousefi_3}, \cite[eq.~(6)]{MecSht}). Since the channel capacity is lower-bounded by $I_{XY}$, this result implies that the capacity grows at least as fast as $(1/2)\log \mathrm{SNR}$, when $\mathrm{SNR} \to \infty$.

\section{Conclusions}
By using a rigorous channel model based on the exact conditional PDF for the soliton amplitudes in (\ref{PDF}), an exact closed-form expression for a lower bound on the capacity of the nonlinear optical fibre channel with no inline dispersion compensation was derived. It has been analytically demonstrated that the lower bound on the capacity for the channel based on the individual amplitudes of well separated solitons displays an unbounded growth similarly to the linear Gaussian channel.

\end{document}